\documentclass[twocolumn,pre,amsmath,amssymb,
aps,floatfix]{revtex4-1}
\usepackage{graphicx}
\usepackage{amsfonts}
\newcommand{\beq}{\begin{equation}}     \newcommand{\eeq}{\end{equation}}
\newcommand{\bal}{\begin{align}}     \newcommand{\eal}{\end{align}}
\newcommand{\beqa}{\begin{eqnarray}}    \newcommand{\eeqa}{\end{eqnarray}}
\newcommand{\bde}{\begin{description}}  \newcommand{\ede}{\end{description}}
\newcommand{\ben}{\begin{enumerate}}    \newcommand{\een}{\end{enumerate}}

\newcommand{\la}{\langle}               
\newcommand{\ra}{\rangle}

\newcommand{\inv}[1]{{\frac{1}{#1}}}

\usepackage{pstricks}
\usepackage{hyperref}
\hypersetup{
colorlinks=true, 
linkcolor=blue,           
}
\begin{document}
\title{
Cellular Fourier analysis for geometrically disordered materials
}
\author{Antoine Fruleux}\email{antoine.fruleux@polytechnique.edu} \affiliation{RDP, Universit\'e de Lyon, ENS de Lyon, UCB Lyon 1, INRAE, CNRS, 69364 Lyon Cedex 07, France} \affiliation{LadHyX, CNRS, Ecole polytechnique, Institut Polytechnique de Paris, 91128 Palaiseau Cedex, France}
\author{Arezki Boudaoud}\email{arezki.boudaoud@polytechnique.edu} \affiliation{RDP, Universit\'e de Lyon, ENS de Lyon, UCB Lyon 1, INRAE, CNRS, 69364 Lyon Cedex 07, France} \affiliation{LadHyX, CNRS, Ecole polytechnique, Institut Polytechnique de Paris, 91128 Palaiseau Cedex, France} 
\date{\today}
\begin{abstract}
Many media are divided into elementary units with irregular shape and size, as exemplified by domains in magnetic materials, bubbles in foams, or cells in biological tissues. Such media are essentially characterized by geometrical disorder of their elementary units, which we term cells. Cells set a reference scale at which are often assessed parameters and fields reflecting material properties and state. Here, we consider the spectral analysis of spatially varying fields. Such analysis is difficult in geometrically disordered media, because space discretization based on standard coordinate systems is not commensurate with the natural discretization into geometrically disordered cells. Indeed, we found that two classical spectral methods, the Fast Fourier Transform and the Graph Fourier transform, fail to reproduce all expected properties of spectra of plane waves and of white noise.  We therefore built a method, which we call Cellular Fourier Transform (CFT), to analyze cell-scale fields, which comprise both discrete fields defined only at cell level and continuous fields smoothed out from their sub-cell variations. Our approach is based on the construction of a discrete operator suited to the disordered geometry and on the computation of its eigenvectors, which, respectively, play the same role as the Laplace operator and sine waves in Euclidean coordinate systems. We show that CFT has the expected behavior for sinusoidal fields and for random fields with long-range correlations. Our approach for spectral analysis is suited to any geometrically disordered material, such as a biological tissue with complex geometry, opening the path to systematic multiscale analyses of material behavior.
\end{abstract}
%\keywords{blabla, bloblo, aps}
\maketitle
%\author{Antoine Fruleux \& Arezki Boudaoud\\
%RDP, Universit\'e de Lyon, ENS de Lyon, UCB Lyon 1, INRAE, CNRS, 69364 Lyon Cedex 07, France; \\
%LadHyX, CNRS, Ecole polytechnique, Institut Polytechnique de Paris, 91128 Palaiseau Cedex, France}

\section*{Introduction}
The past decades have seen a growing interest in geometrically disordered media~\cite{sadoc1999foams} such as liquid and solid foams~\cite{weaire2001physics,weaire1994stress,gibson1999cellular}, granular materials~\cite{slotterback2008correlation}, or biological tissues~\cite{trepat2018mesoscale,paluch2018biophysics}. This brought many questions and concepts related to the dynamics of these media such as coarsening~\cite{fortuna2012growth,hilgenfeldt2001dynamics,mayer2004heterogeneous,lambert2010coarsening,furuta2016close,duplat2011two}, fluctuations~\cite{donev2005unexpected}, jamming transition~\cite{liu2001jamming,majmudar2007jamming,abate2006approach,liu1998jamming,katgert2013jamming}, grain growth~\cite{brechet1999cellular}, or applicability to living tissues~\cite{lecuit2007cell}. Many experimental approaches were developed to observe and quantify cell tilings in these media.  For instance, magnetic resonance imaging~\cite{gonatas1995magnetic} or X-ray tomography \cite{lambert2010coarsening,meagher2011analysis} enable imaging of foam evolution in 3D. Imaging of biological tissue is performed with serial block-face scanning electron microscopy~\cite{denk2004serial} or with confocal microscopy of living samples~\cite{lecuit2007cell}. Using efficient algorithms such as the watershed transform\cite{roerdink2000watershed}, it has been possible to segment these 2D and 3D images, i.e. to extract the geometry and the arrangement of the cells, as performed in foams~\cite{lambert2007experimental}, granular material~\cite{saadatfar2012mapping}, or in biological tissues~\cite{de2015morphographx}. Here, we consider quantitative analyses of properties or fields defined on such segmented images.

\begin{figure}
\centering
\includegraphics[width=.45\textwidth]{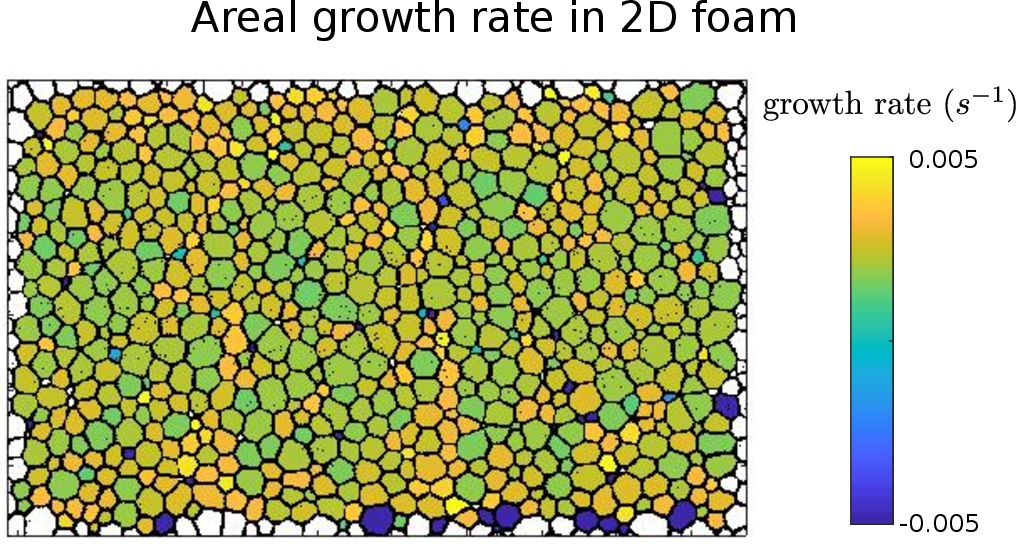}
\caption{Cell areal growth rate in a 2D coarsening foam with visible polydispersity in size. Black lines represent the liquid films between cells, which are colored according to their relative areal growth rate computed over a period of $106$ s (color scale on right). Data courtesy of J\'er\^ome Duplat, see~\cite{duplat2011two}.}
\label{fig:foam}
\end{figure}

An example is given in Figure~\ref{fig:foam}, which shows a coarsening 2D liquid foam. It is disordered, constituted of cells of broadly distributed sizes and irregularly arranged. Accordingly, analyses of fields defined in geometrically disordered media require disentangling potential randomness associated with the field from randomness due to geometry. In addition, the cell often provides a reference scale below which it is difficult, irrelevant, or impossible to define fields. In a foam, areal growth is defined at a discrete level (generally at cell level, Figure~\ref{fig:foam}) because growth requires landmarks (here, vertices) to be computed. These special features make it difficult to assess spatial patterns and test theoretical predictions based on continuous models, such as our prediction of long-range spatial correlations for growth fluctuations in biological tissues~\cite{fruleux2019modulation}. Here, we develop an approach to overcome these difficulties, based on harmonic representation of signals defined on cellular media, which enables to properly analyze the spectra of these signals. We term our approach Cellular Fourier Transform (CFT).

Spectral analysis decomposes signals into linear combinations of harmonics~\cite{helffer2013spectral}. Ad hoc harmonics depend on how signals are represented~\cite{tan2018digital}: for a continuous signal in Euclidean space, it is common to use plane waves and to consider the Fourier transform, while discrete signals defined on regular grids are often decomposed into the eigenvectors of circulant matrices, yielding the Fast Fourier Transform (FFT). Initial frameworks for spectral analysis of signals on irregular grids were based on FFT~\cite{bagchi2012nonuniform}. More recent approaches constructed ad hoc harmonics on graphs~\cite{shuman2013emerging}. Harmonics depend on geometry and, in compact metric spaces, they can be defined as eigenfunctions of the Laplace operator~\cite{helffer2013spectral}. This idea has been extensively used in discrete analysis~\cite{zhang2007spectral}, especially to analyze signals on graphs~\cite{chung1997spectral,spielman2007spectral,shuman2013emerging}. A graph may be endowed with an irregular geometry by ascribing a distance to each edge and define Laplace operators that incorporate distances. However, graphs cannot be used to describe geometrically disordered materials, because graphs do not account for the full geometry of unit cells. Discrete Laplace operators have also been defined for triangular meshing of surfaces~\cite{reuter2009discrete} but their use for signals defined on geometrically disordered materials seems problematic for several reasons. These operators converge only weakly to the smooth Laplace-Beltrami operator in the limit of small mesh size~\cite{rosenberg1997laplacian} and discrete Laplacians on triangular meshes cannot satisfy all desired natural properties~\cite{wardetzky2007discrete}. Other reasons are related to the nature of the mathematical object that we consider: Cellular tessellations of space (2D or 3D), in which cells may have complex shapes and varying topologies (number of neighbors). Therefore the arrangement of cells cannot be encapsulated in binary relations as in triangular meshes or in graphs. 

We therefore built a framework to analyze signals defined on (possibly disordered) tessellations of space. We first motivate our study by applying the Fast Fourier Transform and the Graph Fourier Transform to examples of cellularized signals. We present the geometry of the medium and how signals are represented. We define a coarse Laplace operator, applicable to signals with variations at sub- and supra- cell scale; we show that sine waves are its eigenfunction in the Euclidean space. We project this operator on the cellularized geometry and discretize it. Finally, we test our analysis with numerically generated data, illustrate it with experimental data from a coarsening foam, and discuss the potential applications of this framework. For simplicity, we present results for polygonal tilings of the Euclidean plane, but this method is broadly applicable to domains of any geometry and dimension. We also provide MATLAB (Mathworks) scripts implementing CFT that can be readily adapted to analyze any type of cellularized signal.

\section{Motivation}
We first discuss the applicability of the Fast Fourier Transform (FFT) or the Graph Fourier Transform (GFT) to cellularized signals.

 \begin{figure*}
\centering
\includegraphics[width=\textwidth]{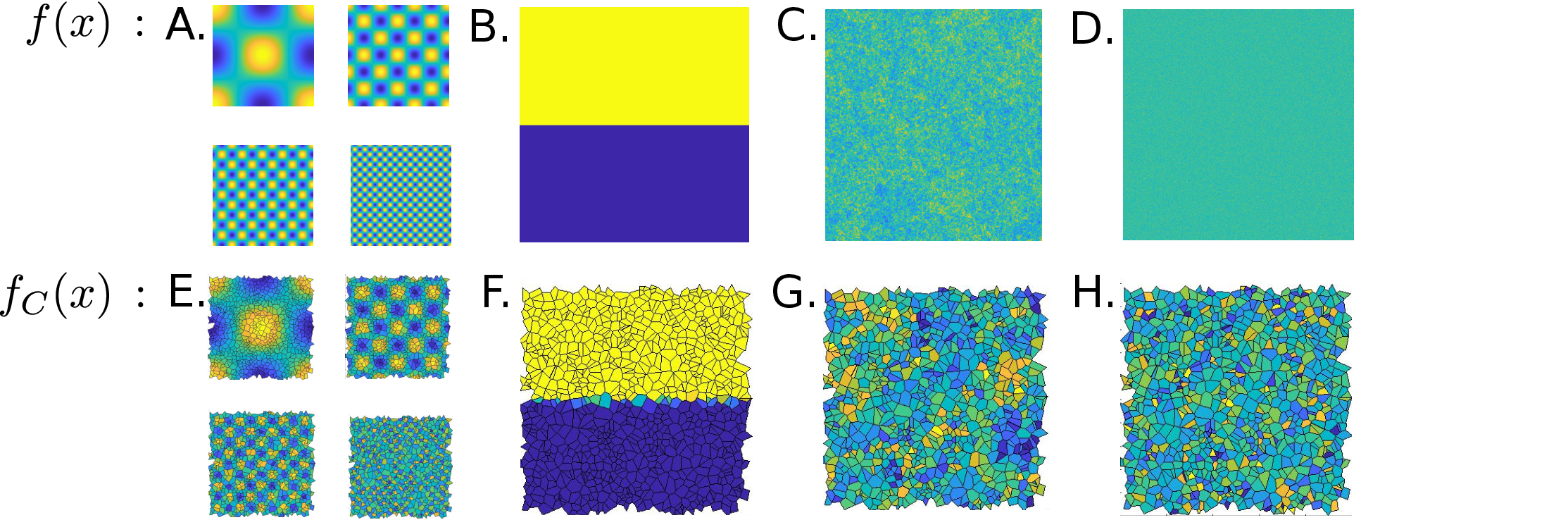} 
\caption{Reference fields: 4 types of continuous fields and their cellularized versions defined on a reference domain. {\bf A}, {\bf B}, {\bf C} and {\bf D} respectively show  stationary waves, a step function, a random field with long range correlation, and a white noise, while {\bf E}, {\bf F}, {\bf G}, and {\bf H} show their respective cellularized versions, { in which each cell is ascribed the average of the continuous field over that cell.}
}
\label{fig:representation}
\end{figure*}

\subsection {Reference tessellation and reference fields}
To test the different approaches, we generated a cellular tessellation and reference fields that associate to each cell a scalar value.  We started from $1000\times 1000$ images of a range of continuous fields shown in Fig.~\ref{fig:representation} {\bf A-D}: plane waves of fixed amplitude and 4 values of wavenumber, step function, long-range correlated noise, and white noise. Stationary waves (Fig.~\ref{fig:representation}{\bf.A}) have wavenumbers $2\sqrt{2}\pi n/ a$ with $n=1,\,3,\,5$ and $10$. The step function is defined as $f(x,y)=-1$ if $y<0$ and $f(x,y)=1$ if $y \geq 0$ (Fig.~\ref{fig:representation}{\bf B}). We built the long-range correlated random field (Fig.~\ref{fig:representation}{\bf C}) using the Fourier filtering method detailed in~\cite{makse1996method}; the 2 points correlation function of pixel intensity $\la I_i I_j\ra$ decays with the distance $d_{ij}$ between pixels like $1/d_{ij}$. The white noise has amplitude 1 (Fig.~\ref{fig:representation}{\bf D}). 

We built a  reference domain and its partition into cells by generating the Voronoi tessellation of a random distribution of 1000 seeds initially placed in a square of side $1000$. We only kept the $894$ cells entirely included in the square, yielding the domain $\Omega$ as shown in Fig.~\ref{fig:representation}{\bf E-H}. $\Omega$ is not a perfect square; { it is a polygon with a large number of edges.} The number of cell neighbors is broadly distributed around 6. { For a given continuous field, we computed the average of the field over each cell; the mapping between a cell and the average field over that cell defines the cellularized field.} The 4 types of cellularized fields are shown in Fig.~\ref{fig:representation} {\bf E-H} and appear as piecewise constant versions of the continuous fields. In the remainder of this section, we focus on stationary waves and on white noise (Fig.~\ref{fig:representation} {\bf A}, {\bf D}, {\bf E} and {\bf H}); we use the step function and the long-range correlated noise in Section~\ref{sec:fourier}.

\subsection{Applicability of Fast Fourier Transform}
We first applied the Fast Fourier Transform (FFT) to reference fields. We discretized these fields on $1000\times 1000$ images, setting to zero the intensity of pixels not belonging to cells (due to the Voronoi tessellation not filling the square). We show the spectra of the reference fields as a function of the $x$ and $y$ component of the wavevector (Fig.~\ref{fig:comparatif} {\bf A} for stationary waves and Fig.~\ref{fig:comparatif} {\bf B} for  white noise) and as a function of the wavenumber in {\bf C}. We made wavevectors  dimensionless using the characteristic cell size $l_c$, defined as the square root of the mean cell area. To ease readability of Figures, we only plotted the first few thousands of modes among the million of possible modes (the starting data has dimensions $1000\times 1000$).

The spectra of the stationary waves show four peaks $(\pm q_x^{(m)},\pm q_y^{(m)})$, lines joining them, and background noise (Fig.~\ref{fig:comparatif}{\bf A}). The four peaks correspond to the wavenumbers of the initial stationary wave while the lines result from a windowing effect due to the Voronoi tesselation not entirely filling the square. The spectrum of the white noise is maximal for the wavevector $(0,0)$ and decays slowly away from $(0,0)$ (Fig.~\ref{fig:comparatif}{\bf B}). In all cases, the amplitude of the FFT increasingly departs from expectations as the wavenumber increases: the height of the peaks decrease for waves of increasing wavenumbers (in Fig.~\ref{fig:comparatif} {\bf C}), instead of having a fixed amplitude, and the spectrum of the white noise decreases with wavenumber (in Fig.~\ref{fig:comparatif} {\bf D}) instead of having a constant amplitude, a phenomenon that is very clear when averaging over realizations of the noise (inset of Fig.~\ref{fig:comparatif} {\bf D}). 

The higher the wavenumber, the less the signal can be approximated by its cellularized projection. Cells act like a low-pass filter that reduces the amplitude of higher FFT modes with respect to the spectra of the initial signals. As we will see later, the method that we developed circumvents this issue. 
 
 \subsection{Applicability of Graph Fourier Transform}
 
The Graph Fourier Transform (GFT) is applicable to signals that are defined on graphs; the harmonics are then the eigenmodes of a discrete Laplace operator. Here a graph can be defined from cells and neigborhood relations between cells: each vertex of the graph corresponds to one cell, and this vertex is connected to all vertices corresponding to neighboring cells. We may account for the metrics of the cellular tesselation by associating to each edge (link between graph vertices $i$ and $j$) a weight $w_{ij}$ that depends on the distance between cell centers $d_{ij}$, $w_{ij}=f(d_{ij})$; { the weight $w_{ij}$ vanishes if the two cells are not neighbours.} For a field $\phi$ that has values $\phi_i$ for $i$ spanning vertices, the value of the Laplacian at vertex $i$ is $(\mathcal{L}(\phi))_i=\sum_{j}w_{ij}(\phi_j-\phi)$. For a fair comparison with the method that we developed, we used the same kernel $f(r)=\exp(-r/\sigma)$ with $\sigma=7 l_c$. We computed the eigenvectors $e_k$ of  the Laplacian $\mathcal{L}$, which are associated to the eigenvalues $\hat{L}_k$. The Graph Fourier Transform of a field is then given by the components of the field with respect the basis $\{e_k\}$.
 
\begin{figure}
\centering
\includegraphics[width=.47\textwidth]{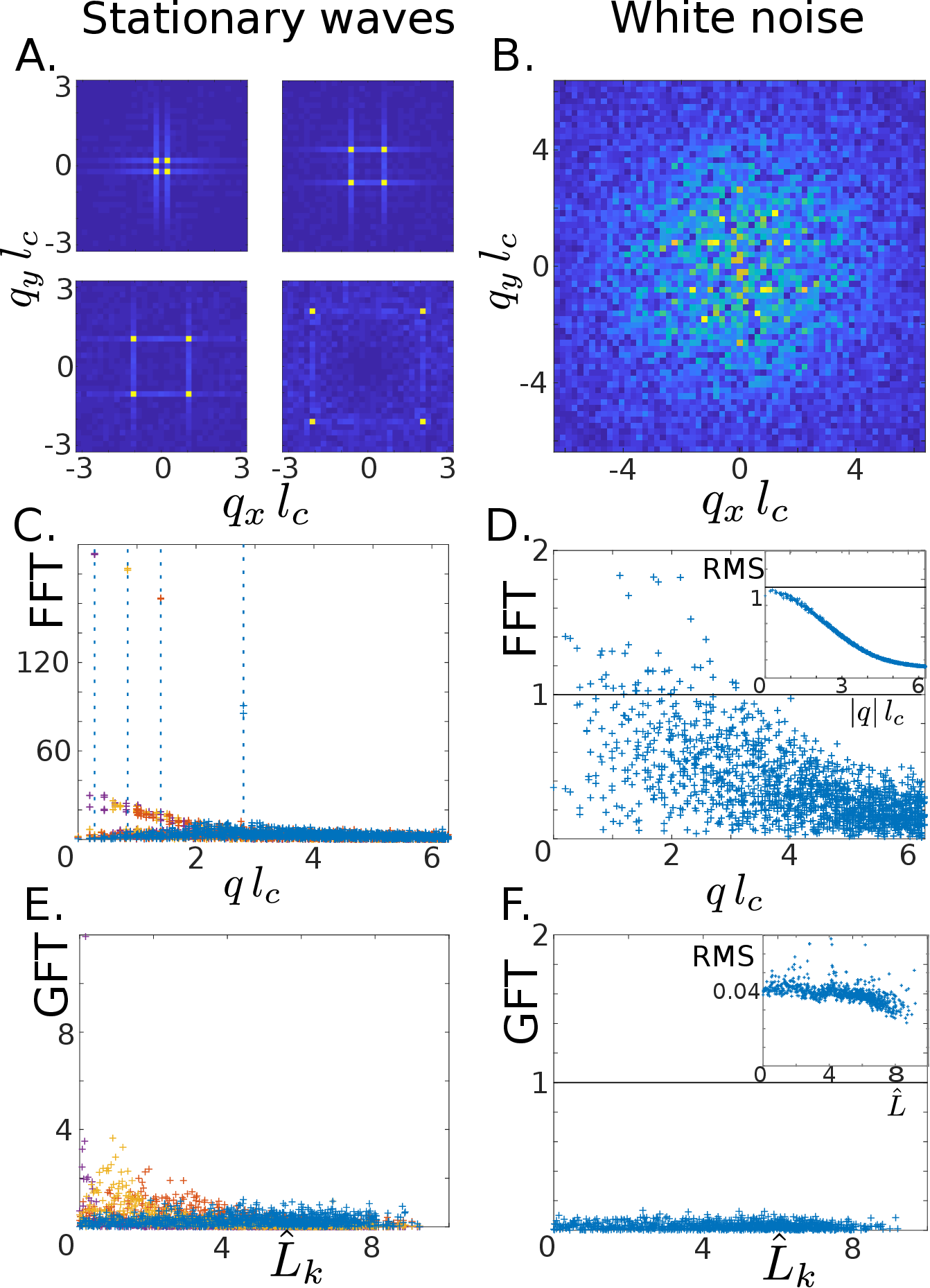} 
\caption{ Fast Fourier Transform and Graph Fourier Transform applied to cellular representations of plane waves and white noise (Fig.~\ref{fig:representation} {\bf E} and {\bf H}). FFT modulus is plotted as a function of the dimensionless components of wavevector $(q_x,q_y)$ in {\bf A} and {\bf B} and as function of wavenumber $q=\sqrt{q_x^2+q_y^2}$ in {\bf C} and {\bf D}, using mean cell size $l_c$ as unit length. In {\bf E} and {\bf F}, GFT modulus is plotted as a function associated eigenvalue $\hat{L}_k$ of Laplacian.  Insets in {\bf D} and {\bf F} show the root mean square spectra obtained by averaging over $1000$ realizations of the white noise. Dashed vertical lines in {\bf C} are at the wavenumbers of the initial fields, while dark horizontal lines in {\bf D} and {\bf F} correspond to the spectral density of the initial noise (theoretical curve).
}
\label{fig:comparatif}
\end{figure}

Because there is no standard definition of the wavenumber for GFT harmonics, we plotted the spectra as a function { of the eigenvalue $\hat{L}_k$ of  the graph Laplacian $\mathcal{L}$}(Fig.~\ref{fig:comparatif} {\bf E-F}). The GFT behaves poorly for stationary waves, having a marked peak only for the wave with highest wavelength. The GFT spectrum of the white noise has amplitudes that are one order of magnitude lower than expected; in addition, when the spectrum is averaged over realizations of the noise, a slowly decreasing trend appears in the spectrum, in contradiction with { expected constant amplitude}.

The GFT is not really applicable to cellularized fields because it behaves poorly with respect to spectra of the continuous fields and it is difficult to define wavenumbers. An explanation is that geometrical information on cell shape is { partially} lost when considering a weight that depends on distance between cell centers. As we will see later, the method that we developed fully accounts for cell geometry.

\section{Formulation}
We first present the mathematical basis of the Cellular Fourier Transform (CFT). The reader may skip to Section~\ref{sec:implementation} to find the implementation and validation of the CFT. { The main results are given by Eqs.~(\ref{eq:Lij}-\ref{eq:spectrum}) that, together with Eq.~(\ref{eq:TFw}), define the harmonics and their wavenumbers from an appropriate discrete Laplace operator.}

\subsection{Signal representation on a cellularized space}
\label{eq:FunctionRepresentation}
In this section we explain how signals are represented in the cellularized space and we specify desired properties of the harmonics.

We consider a bounded domain $\Omega$ of the $n$-dimensional space $\mathbb{R}^n$, divided into $N$ subdomains $\{\omega_i, i= 0, 1, \dots, N-1\}$, which we call cells. We consider a measure defined on $\mathbb{R}^n$, such as area in $\mathbb{R}^2$. Let $\mu$ and $\mu_i$ be the measures (e.g. areas) of domain $\Omega$ and  cells $\omega_i$, respectively: $ \mu=\sum_{i=0}^{N-1} \mu_i$. We aim at analyzing a signal $f$ defined over $\Omega$ (and assumed to have sufficient regularity for all formulations to be mathematically well-posed). 
{ The smoothed version of the signal $f$ is given by a cell-wise constant function $f_C$. $f_C$ associates each cell $\omega_i$ with the average of the continuous signal $f$ over that cell, $\inv{\mu_i}\int_{\omega_i}d\mu(y)f(y),$ where the integral over cell $\omega_i$ is computed according to the measure $d\mu(y)$. For a Eucledian metric and coordinates $(y_1,y_2,\dots y_n)$, we have $d\mu(y)=dy_1dy_2.\dots dy_n$.}
We call $f_C$ the representation of $f$. Figure~\ref{fig:representation} shows examples of signals $f$ and their representations $f_C$ on a random Voronoi tessellation. The representation $f_C$ belongs to a vectorial space $\mathcal{E}$ of dimension $N$ (the number of cells $\omega_i$), which we call the representation space. A basis of $\mathcal{E}$ is the set of functions $\{\psi_i\}$ that vanish outside $\omega_i$ and are defined by
\beq
\begin{array}{rl}
\psi_i:& \Omega\rightarrow\mathbb{R}
\\
& x \rightarrow \psi_i(x)=\left\{\begin{array}{ccc}1/\sqrt{\mu_i} & \mathrm{if} & x\in\omega_i
\\
0 &\mathrm{else}&
\end{array}\right.
\end{array}
\eeq
Given the standard scalar product $\la f\cdot g\ra=\int_{\Omega}d\mu(x)f(x)g(x)$ of two functions $f$ and $g$, the basis of functions $\{\psi_i\}$ is orthonormal,
$
\la \psi_i\cdot \psi_j\ra=\delta_{ij},
$
where $\delta_{ij}$ is the Kronecker symbol ($\delta_{ij}=1$ if $i=j$, else  $\delta_{ij}=0$). The representation $f_C$ of $f$ is also its orthogonal projection on the representation space $\mathcal{E}$,
{
\beq
\label{eq:fCproj}
f_C(x)=\sum_{i=0}^{N-1} f_i\, \psi_i(x)\textrm{, with } f_i=\la\psi_i\cdot f\ra.
\eeq
}

{ We note that $f_i=(1/\sqrt{\mu_i})\int_{\omega_i}d\mu(y)f(y)$ is not the signal averaged over the domain $\omega_i$. The prefactor $\sqrt{\mu_i}$ is important for the present analysis to be mathematical sound and the definition of $f_i$ is the appropriate discretization of the cellularized signal. Inappropriate discretization partly explains the poor results of GFT.} In the following, we aim to define another orthogonal basis for the representation space $\mathcal{E}$ so that its elements enable spectral analysis in cellular media. We will call the elements $e_k$ of this basis harmonics of the representation space. They can be written as
{ $e_k(x)=\sum_i U_{ki}\psi_i(x)$},
where $U_{ki}=\la e_k\cdot\psi_i\ra$ are the elements of the rotation matrix $U$ between the two bases. Finding the harmonics of the representation space is equivalent to { determining} the unitary matrix $U$.

\subsection{Coarse Laplace operator}
\label{sec:IOp&Fourier}
In infinite Euclidean space, Fourier harmonics are plane waves. These plane waves are notably eigenfunctions of the classical Laplace operator. More generally, these plane waves are eigenfunctions of all integral operators that are invariant by translation, a property that we will use to define the harmonics $e_k$. In this section we consider { a Laplace-like integral operator} and investigate its properties, first in infinite Euclidean space and then in bounded domain; we then explain how the problem can be discretized to define the harmonics $e_k$.

\subsubsection{In unbounded space}

Harmonics are often defined as eigenfunctions of the Laplace operator. Because the signals that we consider are smoothed out of their subcellular variations, we build a coarse version of the Laplace operator, formulated as an integral operator $\mathcal{L}$ which, to each function $f$ defined on $\mathbb{R}^n$, associates
\beq
\mathcal{L}\left[f\right](x)=\int_{\mathbb{R}^n}w(|x-z|)(f(x)-f(z))d\mu(z).
\eeq
The kernel $w$ is an integrable real function and $|x-z|$ is the Euclidean distance between points $x$ and $z$ of $\mathbb{R}^n$. We assume that $w(r)$  has a maximum at $r=0$ and vanishes when $r\to\infty$, with a characteristic decay length $\sigma$. Like the discrete Laplacian on a grid, {the operator $\mathcal W$} averages the difference between the local field $f(x)$ and its value on the neighborhood of $x$. In the limit where the lengthscale $\sigma$ vanishes, $\mathcal{L}\left[f\right]\simeq C \nabla^2f$, where $\nabla^2$ is the classical Laplace operator and the constant $C=-\int_{\mathbb{R}^n}w(|z|)\,z^2/2\, d\mu(z)$. The operator $\mathcal{L}$ can be seen as a coarse version of the { Laplacian $\nabla^2$, see} \cite{Note1}.

In all generality, plane waves of wavenumber $q$, { $u_q(x)=\exp(I q\cdot x)$}, (with $I^2=-1$) are eigenfunctions of $\mathcal{L}$: {
 \begin{align}
 \label{eq:TFw}
&\mathcal{L}[u_q](x)= \hat{L}(|q|)u_q(x)\textrm{, with}\\
 &\hat{L}(|{q}|)=\hat{W}(0)-\hat{W}(|{q}|),\nonumber\\
&\hat{W}(|{q}|)=\int_0^{+\infty} w(r)A_d(|q|r)r^{d-1}dr \textrm{, and\nonumber}\\
& A_n(r)=\frac{2 \pi^{(n-1)/2}}{\Gamma((n-1)/2)}\int_0^\pi d\theta (\sin \theta)^{n-2} \exp(I |q| r\cos \theta).\nonumber
\end{align}
}
We will later consider the case when $\hat{L}$ is positive and monotonously increasing on $[0,+\infty[$ so that the eigenfunctions of $\mathcal{L}$ associated to a given eigenvalue are linear combinations of plane waves with all wavenumber having the norm $|q|$.

\subsubsection{In bounded domain}
\label{sec:bounded}
We now consider a compact bounded domain $\Omega$ of $\mathbb{R}^n$ and functions $f$ defined on $\Omega$, { and we replace the operator of the previous section by} 
\beq
\label{eq:gW}
\mathcal{L}\left[f\right](x)=\int_{\Omega}w(|x-y|)(f(x)-f(y))\,d\mu(y),
\eeq
$\mathcal{L}$ is a compact and self-adjoint operator; according to the spectral theorem, the eigenfunctions of $\mathcal{L}$ form an orthogonal basis of the associated Hilbert space~\cite{helffer2013spectral}. If $w$ is well localized around $0$ (equivalently if its decay length $\sigma$ is small with respect to size of { the} domain $\Omega$), then the integral in (\ref{eq:gW}) can be approximated by an integral over the whole Euclidean space $\mathbb{R}^n$ provided that the position $x$ is not too close to the boundary of $\Omega$.  Therefore, Equation~(\ref{eq:TFw}) implies that the eigenvalues of $\mathcal{L}$ are approximately $\hat{L}(|{q}|)$. Associated eigenfunctions are locally approximated by linear combinations of plane waves with wavenumbers of norm $|q|$, except close to the boundary; eigenfunctions may show boundary layers of width $\sigma$ (see below). Note that, in order to reduce this boundary effect for eigenfunctions associated to the lowest eigenvalues, we defined $\mathcal{L}[f](x)$ in Eq.~(\ref{eq:gW}) using the difference $f(x)-f(y)$, which enforces constant functions to be eigenfunctions of $\mathcal{L}$ { that are} associated to the eigenvalue $0$.

\subsection{Harmonics of the representation space}
\label{sec:repspace}
In this section we discretize the problem and consider the representation of the coarse Laplace operator, { $\mathcal{L}$,} defined above. We introduce in section \ref{sec:discreteOp} the harmonics of the representation space and investigate in section \ref{sec:eigenvalues} how they are related to the eigenfunctions of $\mathcal{L}$ and how their wavenumber can be computed. In section \ref{sec:BC}, we introduce a correction to reduce boundary effects.

\subsubsection{Discrete Laplace operator}
\label{sec:discreteOp}
In order to build the representation of the coarse Laplace operator, we first consider the representation $w_C(x,y)$ { of the kernel $w(x,y)$ assumed to be a function of the distance $|x-y|$ so that $w(x,y)=w(|x-y|)$.
$w(x,y)$ is a function of two variables and its representation is piecewise constant on $\Omega^2$. Accordingly, we generalize the representation of a function of a single variable defined as a projection on $\mathcal{E}$ and define the projection of $w(x,y)$ on $\mathcal{E}^2$ as follows: 
\beq
 \label{eq:WC}
 w_C(x,y)=\sum_{i=0}^{N-1}\sum_{j=0}^{N-1}W_{ij}\,\psi_i(x)\psi_j(y),
 \eeq
  where the elements of the weight matrix $W$ are given by}
\beq
W_{ij}=\inv{\sqrt{\mu_i\mu_j}}\int_ {\omega_i}\int_{\omega_j} w(|x-y|)d\mu(x)d\mu(y).
\eeq
The integral operator $\mathcal{L}_C$ associated to the kernel $w_C$ is,
$
\mathcal{L}_C\left[f\right](x)=\int_{\Omega}w_C(x,y)(f(x)-f(y))d\mu(y)
$.
{ Using Eq.~(\ref{eq:WC}) for $w_C$ and the relations $\int_\Omega \psi_j(y)d\mu(y)=\sqrt{\mu_j}$ and $f_j=\la\psi_j\cdot f\ra$, we rewrite the integral operatror as, 
{\small{ $
\mathcal{L}_C\left[f\right](x)=\sum_{i=0}^{N-1}\sum_{j=0}^{N-1} \psi_i(x) W_{ij}\left(f(x)\sqrt{\mu_j}-f_j\right)
$}}.
We then apply this operator to a function $f_C(x)=\sum_{i=0}^{N-1}f_i\,\psi_i(x)$ of the representation space and use the relation $\psi_i(x)\,\psi_j(x)=(1/\sqrt{\mu_i})\,\delta_{ij}\psi_i(x),$ so as to get
 $
\mathcal{L}_C\left[f_C\right](x)=\sum_{i=0}^{N-1}\sum_{j=0}^{N-1}L_{ij}\,f_j\,\psi_i(x),
$ where,}

\beq
L=D-W,\quad D_{ij}=\delta_{ij}\sum_{k=0}^{N-1} W_{ik}\sqrt{\frac{\mu_k}{\mu_i}}.
\eeq
We call { the matrix} $L$ the discrete Laplace operator. It shows similarities with classical discrete Laplace operators~\cite{wardetzky2007discrete}, such as being symmetric, having positive weights $W_{ij}$, and being positive semi-definite. We nevertheless emphasize that these classical discrete Laplacians do not operate on the same mathematical objects: they apply to functions defined on meshes or on graphs, whereas { our operator $L$ applies to functions that are piecewise constant over each cell.} 

{ \subsubsection{ Eigenfunctions and eigenvalues}}
\label{sec:eigenvalues}
The discrete Laplace operator $L$ can be diagonalized and we propose to define the unitary matrix $U$ introduced in section \ref{eq:FunctionRepresentation} from the eigenvectors of $L$, yielding
\beq
\label{eq:vpLij}
L_{ij}=\sum_{k=0}^{N-1} \hat{L}_k\, U_{ki}U_{kj},
\eeq
where $\{\hat{L}_k, k=0, 1, \dots, N-1\}$ are the eigenvalues of $L$. Note that the eigenvector associated to eigenvalue $0$ is $(\sqrt{\mu_0},\sqrt{\mu_1}, \dots, \sqrt{\mu_{N-1}})$ instead of being $(1,1, \dots, 1)$ as for classical discrete Laplace operators. { This difference illustrates how cell size heterogeneity is taken into account by the Cellular Fourier Transform}. The harmonics $\{e_k, k=0, 1, \dots, N-1\}$ associated to $U$ are the eigenfunctions of $\mathcal{L}_C$:
\beq
\label{eq:ekEF}
\mathcal{L}_C[e_k](x)=\hat{L}_k e_k(x).
\eeq

{ In the following, we  show that harmonics of the representation space are good representations of harmonics of the whole space of functions defined on omega and we determine the wavenumbers associated with the eigenfunctions $e_k$, by using the coarse Laplace operator $\mathcal{L}$.  To do so, we first} consider an eigenfunction $f$ of $\mathcal{L}$: $\mathcal{L}[f](x)=\lambda f(x)$. Based on the definitions above, its representation $f_C$ verifies, 
%\beq
%\begin{widetext}
\begin{align}
\label{eq:conv}
&\mathcal{L}_C[f_C](x)=\lambda f_C(x)\\
&+\sum_{i=0}^{N-1}\frac{\psi_i(x)}{\sqrt{\mu_i}}\int_{\omega_i}d\mu(z)\int_{\Omega}d\mu(y)w(|z-y|)(f(y)-f_C(y)).\nonumber
\end{align}
%\end{widetext}
%\eeq 
In the integral in (\ref{eq:conv}), $f(y)-f_C(y)$ varies quickly, {\it i.e.} at the typical cell scale $l_c=(\mu/N)^{1/n}$. If the decay length $\sigma$ of the kernel $w$ is greater than or comparable to $l_c$, then the integral in (\ref{eq:conv}),  { yielding 
$
\mathcal{L}_C[f_C](x)=\lambda f_C(x),
\label{eq:ekEF2}
$
and $f_C$ is, within a good approximation, an eigenfunction of $\mathcal{L}_C$, associated to the same eigenvalue $\lambda$. Therefore the linear space generated by the eigenfunctions of $\mathcal{L}$ associated with eigenvalue $\hat{L}(|q_k|)$ (see Eq.~\ref{eq:TFw})) is projected to the linear space generated by the harmonics $e_k$ such that 
\beq
\label{eq:vp}
\hat{L}_k=\hat{L}(q_k).
\eeq
Finally, based on section \ref{sec:IOp&Fourier}, each eigenfunction of $\mathcal{L}$ associated to eigenvalue $\hat{L}(|q_k|)$ is locally well approximated by a linear combination of plane waves with wavevectors having the same norm $|q_k|$. Therefore the $e_k$ are appropriate harmonics of the representation space because they are the projections of linear combinations of plane waves of the same wavenumber $q_k$, defined by Eq.~(\ref{eq:vp}).}

\subsubsection{Correcting boundary effects}
\label{sec:BC}
The results obtained in Sections~\ref{sec:repspace}-\ref{sec:eigenvalues} are only valid in the bulk of domain $\Omega$. Indeed, { when we explicitly computes the set $\{e_k, k= 0, 1, \dots, N-1\}$ following the above, we found} that some of the harmonics $e_k$ show visible boundary layers near the edges of $\Omega$. This can be explained qualitatively as follows: At edges, $\mathcal{L}[f](x)$ is an integral over a domain about twice smaller than in the bulk; consequently, variations of eigenfunctions at edges are about twice as big as in bulk. Two qualitatively similar methods could be used to correct this artifact. One is to rescale the weight function $w(|x-y|)$ by its integral over $\Omega$, {\it i.e.:} use the kernel $w(|x-y|)/\int_{\Omega}w(|x-z|)d\mu(z)$. Another is to directly rescale each row $i$ of the discrete Laplace operator by $D_{ii}=\inv{\mu_i}\int_{\omega_i}\int_{\Omega}w(|x-y|)d\mu(y)d\mu(x)$. Here we use the latter method and we consider the rescaled Laplace $\bar{L}$ operator,

%\beq
\begin{multline}
\label{eq:Lij}
\bar{L}_{ij}=\delta_{ij}-\bar{W}_{ij},\,\,\, \mathrm{with,}\\ \bar{W}_{ij}=\sqrt{\frac{\mu_i}{\mu_j}}\frac{\int_ {\omega_i}\int_{\omega_j} w(|x-y|)d\mu(x)d\mu(y)}{\int_ {\omega_i}\int_{\Omega} w(|z-t|)d\mu(z)d\mu(t)}.
\end{multline}
%\eeq

 This rescaling only marginally changes the matrix $L$, though it breaks its symmetry. For this reason $U$ is no longer defined as the eigenvectors of $\bar{L}$ but as its right-singular vectors. This does not significantly affect $U$, except for columns corresponding to cells at the edges of $\Omega$. The singular values $\hat{L}_k$ of the rescaled Laplace operator are therefore still related to wavenumbers via  Eq.~(\ref{eq:vp}). If $V$ are the left-singular vectors, we may write the components of $\bar{L}$ as,
 \beq
 \label{eq:Lij2}
 \bar{L}_{ij}=\sum_{k=0}^{N-1}\hat{L}_k\,V_{ki} U_{kj},
 \eeq
 { The wavenumber $q_k$ can be defined from the eigenvalue $\hat{L}_k$ using Eq.~(\ref{eq:TFw}),}
 \beq
 \label{eq:Lij3}
 \hat{L}_k=\hat{L}(q_k)/\hat{L}(0),
 \eeq
 From this analysis, we can deduce the spatial spectrum of a signal $f$. The $k$-th spectral coefficient is given by
\beq
 \label{eq:spectrum}
\hat{\Phi}_k=\la e_k\cdot f \ra=\sum_{i=0}^{N-1} U_{ki}\inv{\sqrt{\mu_i}}\int_{\omega_i} d\mu(x) f(x),
\eeq
and is associated to the wavenumber $q_k$ { defined in Eq.~(\ref{eq:Lij3})}. We call this spectrum the Cellular Fourier Transform: It is the appropriate equivalent of the classical Fourier spectrum in $\mathbb{R}^n$ for signals defined at cell level. {  The spectrum is fully defined by Eqs.~(\ref{eq:Lij}-\ref{eq:spectrum}) together with  Eq.~(\ref{eq:TFw})}

\section{Implementation and results}
\label{sec:fourier}

\subsection{Implementation}
\label{sec:implementation}
{  We implemented the Cellular Fourier Transform (CFT) on the domain and for the fields shown in Fig.~\ref{fig:representation}}. The CFT relies on defining harmonics as the eigenvectors of the discrete Laplace operator given by Eq.~(\ref{eq:Lij}). This equation involves cell measures $\mu_i$ (e.g cell areas in 2D) and a kernel $w(r)$ that vanishes for large $r$.We tested several kernels $w$ and several values of their characteristic decay lengths $\sigma$ (see below). We numerically computed the integrals in Eq.~(\ref{eq:Lij}) using Gauss quadrature. For tractability of numerical calculations, we used the approximation $w(r>6\sigma)=0$. We obtained the harmonics and associated wavenumbers { (Eqs.~\ref{eq:Lij2}-\ref{eq:Lij3})} using the singular value decomposition algorithm of MATLAB (Mathworks), see Supplementary information. We present in Section~\ref{sec:res} the CFTs of a few  results of our approach before testing in Section~\ref{sec:testW} how the parameters of the kernel influence the analysis. As an illustration, we also performed an analysis of the foam coarsening data shown in Figure~\ref{fig:foam}, which is presented in Section~\ref{sec:illustration} .

\begin{figure}
\centering
\includegraphics[width=.45\textwidth]{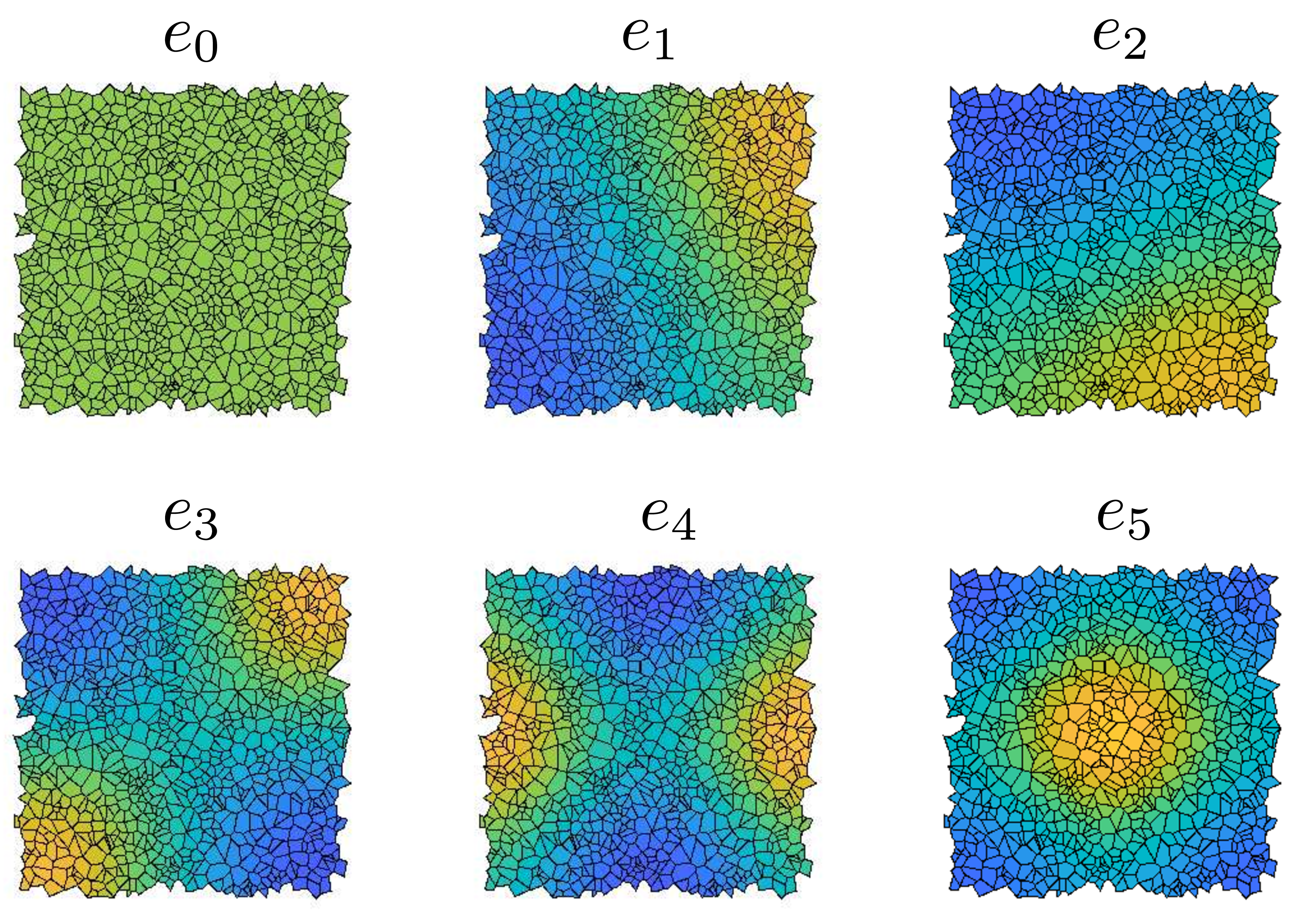} 
\caption{The 6 first harmonics on a domain $\Omega$ that was generated using a randomly seeded Voronoi tessellation. Blue and yellow correspond respectively to negative and positive values of the harmonic. Kernel: exponential, $\sigma=7\,l_c$.}
\label{fig:Fmodes}
\end{figure}

\subsection{CFT applied to artificial fields}
\label{sec:res}

\begin{figure}
\centering
\includegraphics[width=.45\textwidth]{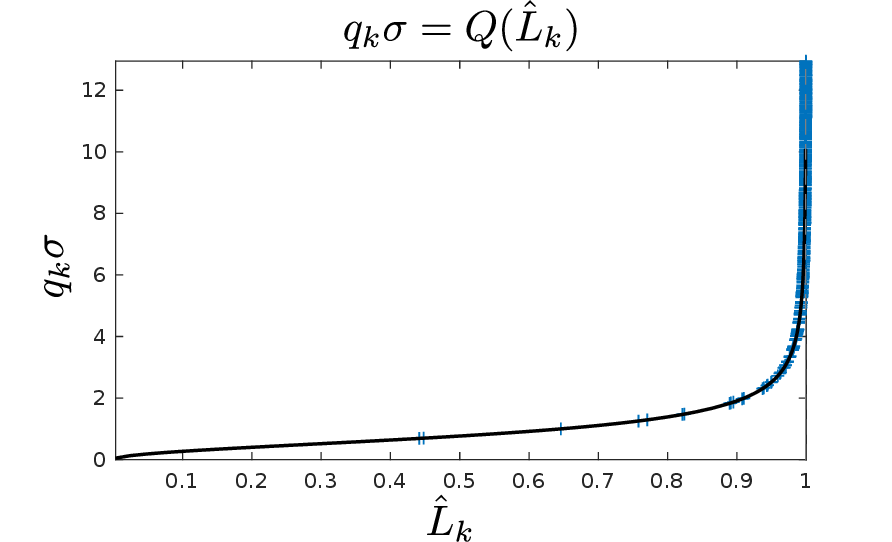} 
\caption{The relation between wavenumbers and eigenvalues of $L$. The black line shows the function $Q$ and the blue crosses show $Q(\hat{L}_k)$ that correspond to the discrete wavenumbers. Kernel: exponential kernel, $\sigma=7\,l_c$}
\label{fig:Q_L}
\end{figure}

In this section, we use an exponential kernel, $w(x)=\sigma^{-2}\exp(-r/\sigma)$. Following (\ref{eq:vp}), the corresponding wavenumbers $q_k$ take values $q_k=1/\sigma \, Q(\hat{L}_k)$, where $Q(l)=\sqrt{(1-l)^{-2/3}-1}$  is the inverse function of $\hat{L}$ introduced in (\ref{eq:TFw}). Note that the wavenumber $q_k$ associated to the harmonics $e_k$ is not given by the square root of $\hat{L}_k$, as would be for the Fourier transform in infinite space (though $q_k\sim \sqrt{\hat{L}_k} $ for small $\hat{L}_k$). These wavenumbers are shown in Figure~\ref{fig:Q_L}.

Figure~\ref{fig:Fmodes} shows the first harmonics $e_k$ on $\Omega$. These harmonics resemble linear combinations of the eigenfunctions of the Laplace operator with Neumann boundary conditions on a square of side $a$~\cite{gottlieb1985eigenvalues}, $u_{m n}(x,y) =\cos (\pi m x/a) \cos (\pi n y/a),$ where $(x,y)$ are the Cartesian coordinates and $(n,m)$ are integers. Indeed,
{ the shape of $\Omega$ is close to a square and in the limit $\sigma\rightarrow 0$, the coarse Laplace operator $\mathcal{L}$ defined in Eq.~(\ref{eq:gW}) converges towards the classical Laplace operator with Neumann conditions on the boundaries of its domain $\Omega$.}~\footnote{{ In the limit $\sigma\rightarrow 0$, the coarse Laplace operator converges toward $ C \nabla^2f$ in the bulk and toward $A(N\cdot \nabla)f$ at the boundaries where $N$ is the unit normal vector oriented inward $\Omega$ and the values of the constants are $C=-{\protect\int}_{\mathbb{R}^n}w(|z|)\,z^2/2\, d\mu(z)$ and $A={\protect\int}_{-\infty}^{+\infty}{\protect\int}_{0}^{+\infty}w(\sqrt{x^2+y^2}) y dxdy$. An eigenfunction $f$ of $\mathcal{L}$ associated to the eigenvalue $\lambda$ verifies therefore $C\nabla^2 f=\lambda$ in the bulk and $A(N\cdot \nabla)f=\lambda$ at the boundaries. Moreover in the same limit $\sigma\rightarrow 0$, the eigenvalue $\lambda$ is negligible at the boundaries, and the eigenfunctions of $\mathcal{L}$ are also the eigenfunctions of the Laplace operator with Neumann boundary conditions on $\Omega$.
}}.
  Accordingly, the first harmonic of the discrete Laplace operator, $e_0$, is constant like $u_{00}$. The following harmonics $e_1$ and $e_2$ correspond respectively to $u_{10}$ and $u_{01}$ respectively but their orientation deviates slightly toward diagonals. More generally, a given harmonic $e_k$  corresponds to a linear combination of eigenfunctions $u_{mn}$ with indices corresponding to $\pi/a\sqrt{m^2+n^2}\sim q_k$. This linear combination is such that higher harmonics $e_k$ tend to have the same spatial periodicity in all directions. { Accordingly, the wavenumbers of the CFT modes are well defined, but the directions of the wavevectors of the modes are not. Indeed, the CFT was designed for generic geometries, which do not have well-defined wavevectors; for instance, classical harmonics in a disk are Bessel functions of the radial coordinate multiplied by sines of the orthoradial coordinate. }

\begin{figure}
\centering
\includegraphics[width=.45\textwidth]{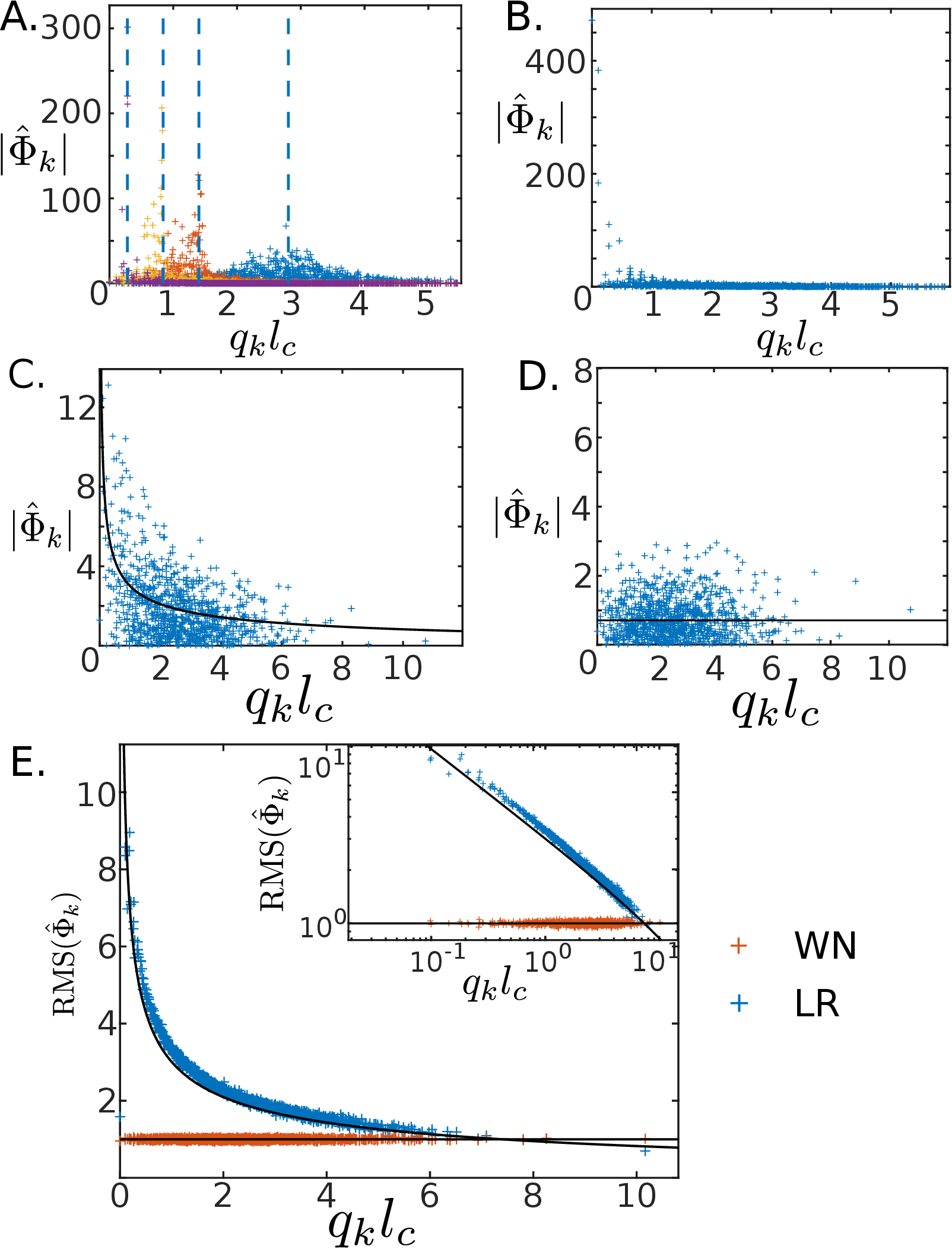} 
\caption{CFT of the cellularized fields shown in Figure~\ref{fig:representation}. {\bf A-D} Spectra of the cellularized fields shown in Figure~\ref{fig:representation}{\bf E-H}, respectively. 
  {\bf E} Root mean square spectra of white noise (red) and of long-range correlated noise (blue); the average is over 1000 realizations 
{ and is shown in linear (main plot) and logarithmic (inset) scales.
The magenta, yellow, red and blue data in {\bf A} are associated to the wavenumbers $2\sqrt{2}\pi n/a$ of the initial continuous signals (represented by dashed lines) with $n=1$, $3$, $5$ and $10$, respectively, and $a$ is the dimension of the square.} The dark lines in {\bf C}, {\bf D} and {\bf E} represent the theoretical curves given by the Fourier transforms of the continuous signals $f$. Kernel: exponential, $\sigma=7\,l_c$.}
\label{fig:tousspectres}
\end{figure}

In Figure~\ref{fig:tousspectres} we show the spectra of the fields plotted in Figure~\ref{fig:representation}.
As expected, the spectra of the stationary waves representations are maximal close to the wavenumbers of the initial continuous waves ((Fig.~\ref{fig:tousspectres}{bf A}). { Nevertheless, the width of the peaks increases with the wavenumber of the initial continuous stationary waves, owing to a less accurate approximation of the stationary wave by its representation in the face of finite cell size. } Also as expected, the spectrum of the step function representation is peaked at 0 (Fig.~\ref{fig:tousspectres}{\bf B}). The spectra of field with long-range correlations (Fig.~\ref{fig:tousspectres}{\bf C}) and of the white noise (Fig.~\ref{fig:tousspectres}{\bf D}) are random and the amplitudes are distributed around zero (here we plot the absolute value); the amplitudes slowly decay with $q_k$ in {\bf C} but stay constant in {\bf D}. To further study random fields, we generated 1000 realizations of the noise. We estimated the average spectrum, which should give an estimate of the Fourier transform of the correlation function. The spectra of representations behave as expected: constant for white noise and power-law decay for field with long-range correlations (Fig.~\ref{fig:tousspectres}{\bf E}). 

{ \subsection{CFT compared to FFT and GFT}}
In the context of cellularized signals, the CFT performs better than the Fast Fourier Transform. They both retrieve peaks in the spectra, but the position of the peaks is less precise with FFT and the peaks are broader for FTT. This can be quantified using the spectral density $\hat{\Phi}_k$; the mean, $q_M$, and root mean square, $q_{RMS}$ of wavenumber are computed as, $(\sum_{k=0}^{N-1}q_k^n|\hat{\Phi}_k|^2)/(\sum_{k=0}^{N-1}|\hat{\Phi}_k|^2)$, with $n=1$ and $n=2$, respectively. We computed the relative shift in wavenumber $\delta=(q_M-q_i)/q_i$ between the resulting spectrum and initial stationary wave. For the CFT, we found values  $\delta=-0.0583$, $-0.127$, $-0.123$, and $-0.0993$ (for the 4 respective stationary waves), which are orders of magnitude smaller than for the CFT, $\delta=3.91$, $1.89$, $1.89$ and $1.76$. Similarly, we computed the relative standard deviation $\nu= (q_{RMS}^2-q_M^2)^{1/2}/q_i$ of the resulting spectrum. For the CFT, we found values $\nu=0.164$, $0.130$, $0.130$, and $0.183$, which are also orders of magnitude lower than for the FFT, $\nu=26.6$, $10.4$, $8.09$ and $5.75$. The CFT retrieves a flat spectrum for white noise, whereas the FFT yields a decaying spectrum. This is due to the FFT adding artificially a large number of degrees of freedom ($\sim10^6$) to those of the cellularized signal (number of cells $\sim10^3$); hence the spectrum of the noise spreads out to larger wavenumbers. Reducing the number of FFT modes would not be a solution because FFT is defined on a square grid which cannot match the topological relations between cells in a disordered medium.

In this context, the CFT also performs better than the Graph Fourier Transform. The GFT yields broad peaks for the spectrum of stationary waves and a decaying spectrum for the white noise. Although this might be improved by a better choice of the weight of links between cells, the GFT would still discard cell shape and the difficulty of defining wavenumbers in the GFT would remain a major issue.

We note, however, that all transforms would yield very similar results for a tessellation with identical square-shaped cells. As such, the CFT is most appropriate for geometrically disordered media.

\subsection{Sensitivity of the CFT to the kernel}
\label{sec:testW}
\begin{figure}
\centering
\includegraphics[width=.45\textwidth]{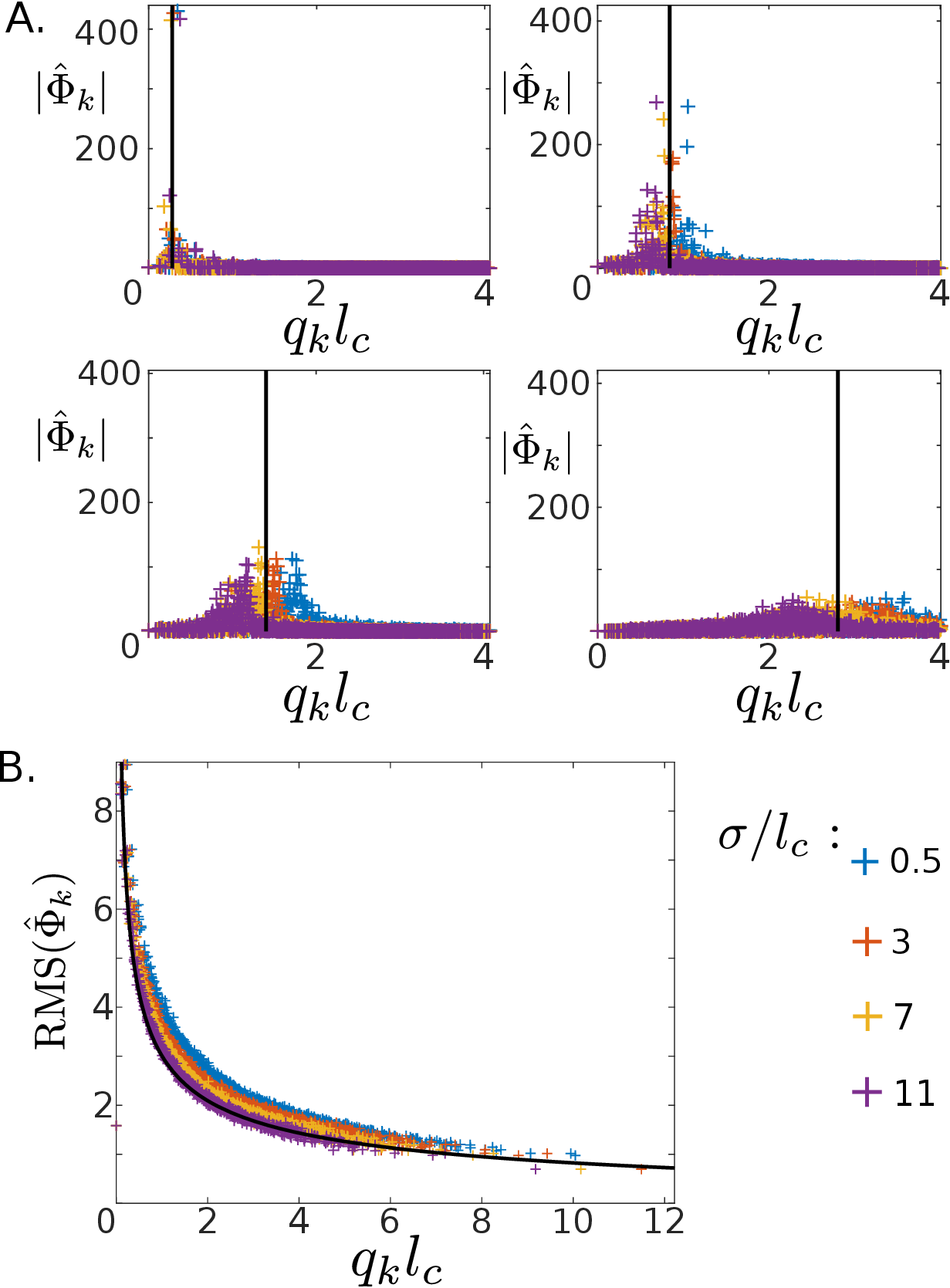} 
\caption{Effect of kernel decay length. Spectra of stationary waves {\bf A} and root mean square spectra of long-range correlated noise {\bf B}. The stationary waves are those represented in Fig.~\ref{fig:representation}{\bf A} and one realization of the random signal was shown in Fig.~\ref{fig:representation}{\bf G}. The different colors correspond to different kernel decay lengths $\sigma$ of the exponential kernel. Black lines represent the Fourier transform of the input signals $f$ (theoretical curves).}
\label{fig:testSigma}
\end{figure}

We first tested the influence of the kernel decay length, $\sigma$, of the exponential kernel. We show in Figure \ref{fig:testSigma} spectra obtained with different values { of $\sigma$}. We see that when $\sigma$ is small, spectra are shifted towards higher wavenumbers. Optimizing the kernel requires taking into account domain size and numerical precision at which the singular value decomposition is performed. The cell size $l_c$ should be a lower bound for $\sigma$ while domain size and numerical precision prevent $\sigma$ from being too large. Optimal values for $\sigma$ must depend on the whole distribution of cell shapes and size as well as on the number of cells. We estimated this optimal value by maximizing the agreement between the wavenumbers corresponding to maxima of spectra in \ref{fig:testSigma}{\bf A.} and the wavenumber of the corresponding stationary waves. We found $\sigma\simeq 7 l_c$, which is intermediate between cell size and domain size.

\begin{figure}
\centering
\includegraphics[width=.4\textwidth]{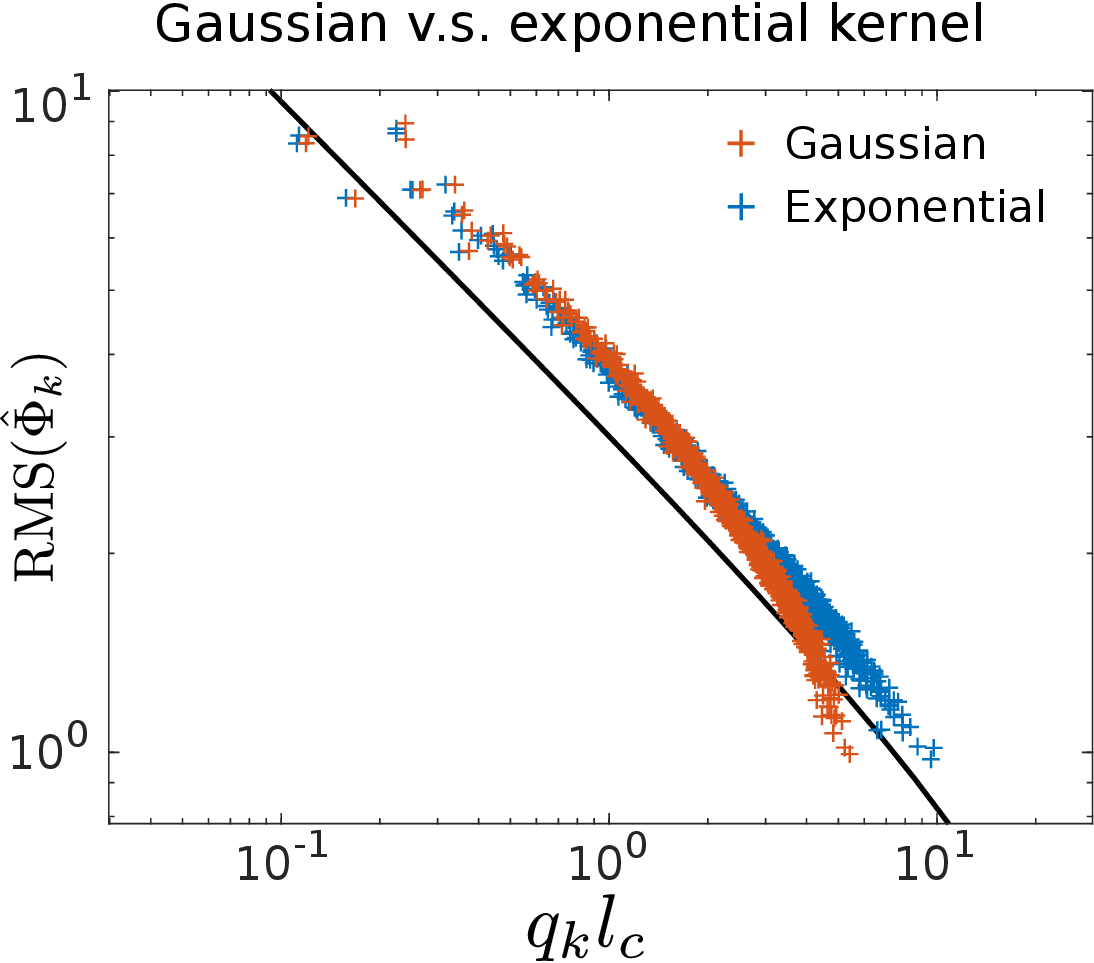} 
\caption{ Effect of kernel type on the spectra. Log-log plot showing the mean square spectra of the representations of fields with long range correlations, as obtained with Gaussian (red) and exponential (blue) kernels ($\sigma= l_c$). Spectra were averaged over 1000 realizations. The black line represents the same root mean square spectra but deriving from the Fourier transform of the input signals $f$ (theoretical curve).}
\label{fig:testKernel}
\end{figure}

We calculated the harmonics of the same domain $\Omega$ using a Gaussian kernel $w(r)=\exp(-{r^2}/({2\sigma}^2))\sigma^{-2}(2\pi)^{-1}$ for which $q_k=1/\sigma Q(L_k)$ with $Q(l)=\sqrt{-2\ln (1-l)}$ and compared the results with those obtained above with an exponential kernel. We did not observe significant differences with the exponential kernel, except at very high frequency. Such differences at high frequency are better seen with long-range correlated fields, as visible in Figure~\ref{fig:testKernel}: The Gaussian kernel leads to an underestimation of the spectrum at large wavenumbers. This can be ascribed to finite numerical precision. With the Gaussian kernel, the eigenvalues $\hat{L}_k$ are distributed closer to $1$ which is a singularity of $Q$. For this reason, a higher precision on $\hat{L}_k$ would be required in the Gaussian case.

\subsection{CFT illustrated with coarsening foam}
\label{sec:illustration}
\begin{figure}
\centering
\includegraphics[width=.35\textwidth]{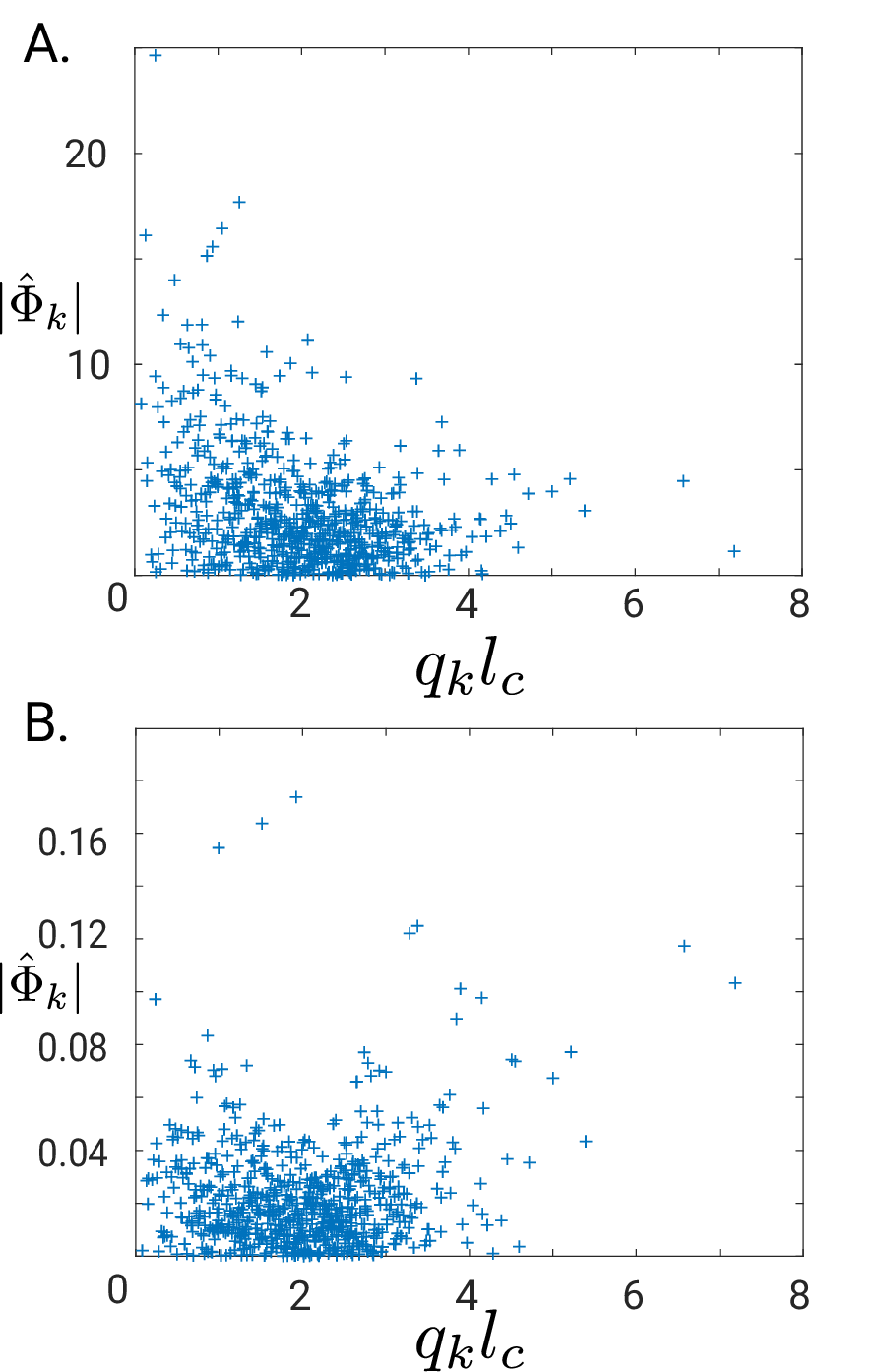} 
\caption{Coarsening foam. Spectrum of areal growth in {\bf A.} and { cell density in {\bf B.}} for the foam shown in Figure~\ref{fig:foam}. Kernel: exponential, $\sigma=7 l_c$.}
\label{fig:sf}
\end{figure}

To illustrate the Cellular Fourier Transform, we analyzed data for a two-dimensional foam shown in Figure~\ref{fig:foam}. We first obtained cell contours using the MATLAB imfill function which is based on morphological reconstruction. We ascribed pixels to cells; we calculated the discrete Laplacian from (\ref{eq:Lij}) by summing over pixels and then computed the harmonics. Figure~\ref{fig:sf} shows the spectrum of areal growth and cell density. The spectrum for growth seems to be random and overall decay with wavenumber. Its resemblance with the spectrum shown in Figure~\ref{fig:tousspectres}{\bf C} suggests long-range correlations for areal growth in foams. { Knowing that, in a coarsening foam, relative areal growth of a cell of area $S$ and number of neighbors $n$ is proportional to $(n-6)/(Se)$ with $e$ average thickness of films, this could be associated with long-range correlations in any of these parameters. The spectrum for cell density $1/S$ is rather flat, with a slight increasing trend at high wavenumbers. This trend would be expected based on a consequence of the Aboav-Weaire and Lewis laws, according to which relatively small cells are be in contact with relatively large cells (and conversely) leading to a short range anticorrelation in cell size~\cite{chiu1995aboav}.} Although this analysis is preliminary { and should be taken with caution,} it indicates that CFT may help reveal new features of geometrically disordered materials.

\section{Conclusion}
{ We considered here the harmonic decomposition of fields defined over disordered cellular media. We found that the classical Fast Fourier Transform and Graph Fourier Transform have some shortcomings in this context. In brief, the FFT is defined on a square grid and does not account for disordered topology (variable number of neighbors), while the GFT does not account for disordered cell shape. We therefore built a more adequate harmonic decomposition}, which we called Cellular Fourier Transform. It is based on the definition of a coarse Laplace operator and the use of an appropriate localized kernel. We found that the resulting harmonics { generally} behave as expected for an exponential kernel of decay length that is intermediate between cell size and domain size, { though the CFT does not yield perfect peaks because of finite cell size. Another limitation of the CFT is that it is not well suited to the identification of wavevectors because it was designed for domains of irregular shapes. Overall,} this harmonic decomposition is suited to disordered media that are divided into cells with variable sizes and irregular arrangements, such as foams, emulsions, granular materials, or biological tissues. As the definition of the harmonics does not depend on the coordinate system, our approach would also be applicable to non-Euclidean geometries, such as curved surfaces embedded in 3D and in particular biological thin tissues with complex 3D shapes.

Our method could be broadly useful for disordered media, even in the absence of subdivisions into cells. In some experimental situations like in fluid dynamics, it is possible to track landmarks such as particles to quantify their displacement. To a certain extent, and similarly to the Helmholtz decomposition of a vector field as the sum of a curl-free field and a divergence-free one, an equivalent point of view would be to consider a triangulation of the landmark distribution and to study the deformation and the rotation of the triangles. One could define invariants which do not depend the translation relative to the reference frame or the coordinate system and directly apply the CFT to those invariants.

Finally, we believe that our approach can be used for cellularized media in all contexts where Fourier transforms are used. This includes statistical estimations (estimating fluctuations at different scales and their correlations), constructing a wavelet decomposition, or pattern recognition. 

\section*{Acknowledgment}
We thank Pierre Borgnat and Chun-Biu Li for fruitful discussions and J\'er\^ome Duplat for sharing raw data of foam coarsening and for useful comments. This work was supported by the Universit\'e de Lyon through the program "Investissements d'Avenir" (ANR-11-IDEX-0007), and by the French National Research Agency through a European ERA-NET Coordinating Action in Plant Sciences (ERA-CAPS) grant (ANR-17-CAPS-0002-01).

\bibliographystyle{unsrt}
\bibliography{biblio}{}
\newpage
\end{document}